# Pressure-Induced Detour of Li$^+$ Transport during Large-Scale Electroplating of Lithium in High-Energy Lithium Metal Pouch Cells

Dianying Liu,[1,†] Bingbin Wu,[1,†] Yaobin Xu,[1] Jacob Ellis,[1] Arthur Baranovskiy,[1] Dongping Lu,[1] Joshua Lochala,[1] Cassidy Anderson,[1] Kevin Baar,[1] Deyang Qu,[2] Jihui Yang,[3] Diego Galvez-Aranda,[4] Katherine-Jaime Lopez,[4] Perla B. Balbuena,[4] Jorge M. Seminario,[4] Jun Liu[*,1,3] and Jie Xiao[*,1,3]

[1]Pacific Northwest National Laboratory, Richland, WA99352, United States
[2]University of Wisconsin Milwaukee, Milwaukee, WI53211, United States
[3]University of Washington, Seattle, WA98122, United States
[4]Texas A&M University, College Station, TX77843, United States
†: These authors contributed equally to this work

Corresponding authors: jie.xiao@pnnl.gov, Jun.liu@pnnl.gov

**Abstract:**

Externally applied pressure impacts the performance of batteries particularly in those undergoing large volume changes, such as lithium metal batteries. In particular, the Li$^+$ electroplating process in large format pouch cells occurs at a larger dimension compared to those in smaller lab-scale cells. A fundamental linkage between external pressure and large format electroplating of Li$^+$ remains missing but yet critically needed to understand the electrochemical behavior of Li$^+$ in practical batteries. Herein, this work utilizes 350 Wh/kg lithium metal pouch cell as a model system to study the electroplating of Li$^+$ ions and the impact of external pressure. The vertically applied uniaxial pressure on the batteries using liquid electrolyte profoundly affects the electroplating process of Li$^+$ which is well reflected by the self-generated pressures in the cell and can be correlated to battery cycling stability. Taking advantage of both constant gap and pressure application, all Li metal pouch cells demonstrated minimum swelling of 6-8% after 300 cycles, comparable to that of state-of-the-art Li-ion batteries. Along the horizontal directions, the pressure distributed across the surface of Li metal pouch cell reveals a unique phenomenon of Li$^+$ migration during the electroplating (charge) process driven by an uneven distribution of external pressure across the large electrode area, leading to a preferred Li plating in the center area of Li metal anode. This work addresses a longstanding question and provides new fundamental insights on large format electrochemical plating of Li which will inspire more innovations and lead to homogeneous deposition of Li to advance rechargeable lithium metal battery technology.

## Introduction

Rechargeable lithium metal batteries have gained significant interest recently due to their potential to double the cell level energy of state-of-the-art lithium-ion batteries (*1*). Great progress has been made on increasing both the cell level energy and cycling stability forward (*2, 3*), due to the fundamental research to discern lithium dendrite formation (*4*), interfacial reactions (*5, 6*), cell failure mechanism under realistic conditions (*7, 8*). This includes the identification of key parameters which include electrolyte amount, cathode mass loading and lithium metal amount (thickness). Together, these dictate the cycling stability of lithium metal coin cells (*5*) and must be carefully tuned to achieve meaningful results during coin cell testing and allow for fair comparisons (*9, 10*).

As testing conditions become more standardized in battery research, results become more reproducible which enables exploration of other key parameter, namely, cell pressure which also profoundly affects the performance of lithium metal batteries (*11, 12*). For all-solid-state batteries, it is known that the externally applied pressure has a significant impact in extending the cycling stability of lithium metal cells and shapes the morphology of repeatedly deposited lithium metal (*13*). For lithium metal cells using liquid electrolyte, there good literature exists which discusses the impact from external pressure while using coin cell batteries (*14, 15*), single layer pouch cells (*16, 17*), or small anode-free pouch cells (*11, 18*). While those exploratory works provide valuable information to deepen the scientific understanding of pressure effects on lithium metal cells, no studies to date look to understand the roles of pressures on practical Li metal batteries with high energy and capacity, leaving key questions unanswered. For example, the size effect - electroplating of lithium on the small area in coin cells is more controllable than the process occurring on a large-format electrodes used in pouch cells. The pressure impacts on electrochemical deposition at an enlarged scale is not well understood but could provide critical information in guiding lithium metal cell design. While pressure plays a role in stabilizing lithium metal cycling it is unknown whether pressure or electrolyte amount play the dominant role in stabilization or the interplay between the two. In order to achieve a high energy density pouch cell, the electrolyte amount is far less than often utilized in literature reports. When lean electrolyte conditions are used, questions remain on how the external pressure impacts both the electrolyte distribution and the utilization of lithium metal within pouch cells.

This work utilizes three identical 350 Wh/kg lithium metal pouch cells to investigate the fundamental relationship between pressure and the electroplating process of lithium metal. Different external pressures are compared and correlated to the performance of lithium metal pouch cells. The pressure distribution on the pouch cell surface is also mapped to understand the reaction activity across the entire surface of the cell upon cycling, providing new insights in designing high-energy lithium metal batteries with a long cycling stability.

## Results and Discussion

There are two common designs for applying external pressures on Li-ion pouch cells (*19*): constant gap or constant pressure. The former fixes a cell between two plates which does not allow outward expansion (Fig. S1A and 1D), while the latter typically adds constant spring force between two plates (Fig. S1B and 1E). For Li-ion batteries, the volume change is relatively limited, thus the "brace" design of constant gap works well. For lithium metal cells, however, the volume change during charge is substantial especially when multiple layers of Li foils are stacked in the same cell. The screws used to fix the two plates in the constant gap design will be slowly pushed up by cell's volume expansion creating additional gap between the two plates during cycling. Therefore, more rigid enclosures are needed to fix the lithium metal pouch cell (*11*). For the constant pressure design, the addition of four springs (Fig. S1B) on top of the two brace plates helps maintain the pressure/force maintained within fixed range depending on spring constant. Through the appropriate selection of springs, the deflection will be minimized, maintaining an almost "constant" pressure on the Li-ion pouch cells. However, due to Li metal batteries' larger volume expansion upon cycling, as compared to Li-ion batteries, the deflection of spring cannot be maintained minimum by simply using the "constant pressure" fixture in Fig. S1B for Li metal pouch cells.

In this work, a modified testing fixture is designed (Fig. S1C and 1F) which combines both constant gap and constant pressure. An additional modified bolt is built in the center of the "constant pressure" brace. This "hybrid" fixture still allows cell to expand and spring to deflect but limits movement through the length of the bolt. A relatively consistent pressure during most of the cycles is ensured, while avoiding significant volume increase at the end of charge. Three Li metal pouch cells are prepared by applying the same design each with a total cell-level energy of 350 Wh/kg (see Table S1, cell design A for details). The three key parameters affecting lithium metal battery cycling stability are kept consistent in the three pouch cells: (1) Li anode thickness (on each side of copper current collector) of 50 μm, (2) $LiNi_{0.6}Mn_{0.2}Co_{0.2}O_2$ (NMC622) cathode with a mass loading of 19.1 mg/cm$^2$ corresponding to 3.5 mAh/cm$^2$ of areal capacity (on each side of the aluminum current collector), and (3) the amount of electrolyte is kept at 1.9 g/Ah in all cells. Without precise control of these three key parameters, it is challenging to develop a conclusive understanding of pressure impacts.

In all pouch cells, the self-generated pressures (blue lines in Fig. 1A-1C) increase upon charge because of plating of additional Li metal layers on the anode side. Note in Li/NMC622 chemistry, Li$^+$ ions are stored in two different locations, both in the NMC cathode and the Li foil anode. Li from NMC cathode is used for plating during charge and stripped back during discharge, while the original 50 μm Li foil functions more like a current collector in the beginning before gradually participating in the electrochemical reactions. The self-generated pressure decreases during discharge when Li is stripped and travels back to the cathode lattice sites. The amplitude of the changes in the self-generated pressure, however, have an inverse relationship with the externally applied pressure on the fresh pouch cells. When 16 and 26 psi are applied, the peak pressures at the end of the first charge reach 140 and 120 psi, respectively (Fig. 1A and 1B). However, when 36 psi is applied on the pouch cell, the highest self-generated pressure is only 70 psi at the end of the first charge (Fig. 1C). Appropriately high external pressure in the beginning compresses the cell and shortens the diffusion length of Li$^+$ ions between two electrodes. Accordingly, the concentration gradient of Li$^+$ established near anode surfaces during plating is also milder in the cell under higher initial pressures promoting the formation of larger Li particles with reduced surface areas (*4*). Therefore, the formation and accumulation of SEI and dead Li is decelerated in the cells under relatively higher initial pressure, which is indirectly reflected by the reduced amplitude of pressure maximum observed during each charge (Fig. 1D-1F). Note that for liquid cells, it is not necessarily true that the higher the external pressure the better. If the external pressure is too high, an internal short (Fig. S2) can be triggered more frequently by the defective sites in the cell, for example, near the tabbing area (*20*). Notably, even the highest pressure of 36 psi used in this work is considerably lower than that used for solid state lithium metal batteries which ranges from 2 MPa to 250 MPa (290 psi to 36259 psi) (*21-23*).

Upon cycling, the pressure at the end of charge increases during almost every cycle (Fig. 1D-1F) in all three pouch cells, indicating the continuous "thickening" of the anode. The pressures at the end of each discharge of pouch cells (Fig. 1D-1F), however, shows almost no change in the first 50-100 cycles (under 16 or 26 psi, Fig. 1D and 1E) or even 250 cycles (under 36 psi, Fig. 1F) and usually goes back to the original applied value. This observation is consistent with the fact that Li being utilized back and forth between the electrodes in the first tens or hundreds of cycles, depending on the external pressure, is mainly from NMC in Li/NMC622 pouch cells. The higher initial pressure reduces the irreversible loss of Li during each cycle because most Li (from NMC) can go back to NMC, an almost fully recover the initial pressure seen at the end of discharge in the first 50-250 of cycles. In the beginning of cycling, Li foil is almost intact and functions as a current collector, while gradually getting involved in the reactions during each cycle, both electrochemically and chemically. This is reflected by the continuous increase of end-of-charge pressure resulting from anode thickening. After extensive cycling, pressures at the end of charge and discharge all increase steadily, particularly towards the end of cycling. Still, the amplitude of pressure increases from the beginning to the end is always less in the pouch cell tested with higher external pressures applied.

In principle, the number of times (cycles) that the end-of-discharge pressure can go back to the original value is an indication of how efficiently Li from NMC side is being utilized. For example, for the pouch cell tested under 16 psi, during the first 1000 hours (Fig. 1D) or 63 cycles, the end-of-discharge pressure always returns close to 16 psi, while for the cell under a pressure of 36 psi, the reversible pressure change lasts for almost 4000 hours (Fig. 1F) or 250 cycles. However, the cycling stability of the three pouch cells under different initial pressures does not display significant differences. For the pouch cell under 16 psi, the cell capacity degrades to 80% of its original capacity after 282 cycles (Fig. 1G). The pouch cells under 26 and 36 psi reach 80% of their initial capacity at 299 (Fig. 1H) and 318 cycles (Fig. 1I) are achieved, respectively. To understand this, one needs to consider the dominant factors that dictate the cycling stability of Li metal batteries. As mentioned earlier, there are a few parameters impacting the cycling of Li metal cells, such as cathode loading, Li thickness and electrolyte amount. From the current results, pressure, although important especially for reproducibility, is not the dominant factor that helps extending the cycling performance of lithium metal batteries in liquid electrolyte. Instead, the amount of electrolyte available in the pouch cell predominantly determines the stable cycling of Li metal batteries. Once the electrolyte dries out, the almost "dried" cell will not respond to the external pressure effectively, leading to the limited cycling variations (Fig. 1G-1I). Another 350 Wh/kg pouch cell with an increased amount of electrolyte of 2.2 g/Ah (compared to 1.9 g/Ah in this work) demonstrated almost 459 stable cycles (Fig. S3 and cell design B in Table S1), confirming that residual electrolyte plays a more critical role than external pressure in extending the cycle life of Li metal batteries. Our previous published work of a 350 Wh/kg pouch cell using 50 μm Li also contained 2.2 g/Ah electrolyte and demonstrated 430 stable cycles (*3*). External pressure helps extending Li metal cycling only when there is still sufficient amount of liquid electrolyte available in the cell.

Another interesting observation is the reduced cell swelling observed in all three pouch cells after extensive cycling (Fig. 2). Before cycling, the pouch cell had a measured thickness of 5.45 mm (Fig. 2A). Even under the lowest pressure of 16 psi, cell swelling is only 8.2% after 304 cycles (Fig. 2B). For the other two pouch cells under 26 and 36 psi external pressures, they both expanded by only 6-7% after more than 300 cycles (Fig. 2C-2D). An appropriately applied external pressure, combined with a balanced cell design, effectively suppressed aggressive volume expansion of Li metal anode during cycling. The cycled lithium metal anodes harvested from three pouch cells are further compared in Figure 2E-2G. Prior to disassembly for characterizations, the pouch cells were in a discharge status meaning the majority of Li from NMC side resided in the cathode. Figure 2E-2J mainly characterizes Li foil after cycling. Intact dense Li is found in all three cycled pouch cells. For consistency, Li taken for characterization is from the same location in three cells as indicated in Figure 2K. Some of those unreacted Li form column-like structures extending from current collector to the surfaces of the Li anode, suggesting the entire pillar did not react at all (Fig. 2E-2G). Those unused Li columns are commonly present in the anode side in all three cycled pouch cells (see surface views in Fig. 2H-2J) and spread within the entire Li electrode samples which will be discussed in a later section (Fig. 3). The formation of these Li columns is related with the utilization of Li from Li foil side. As discussed earlier, initially the Li utilized during cycling is mainly originating from the NMC cathode. Once the irreversible loss from this portion of Li (from NMC) occurs, Li from Li foil side will compensate for that loss and thus participate in the subsequent reaction. Since it is not necessary that all the lithium metal foil needs to be used, intact Li is always left behind in the cycled Li metal anode (Fig. 2E-2J). The observation of completely intact Li "columns" is presumably assigned to the electrolyte wetting which is likely uneven from the beginning. If there are bare regions with no electrolyte coverage on certain locations of Li foils, the reaction will be very limited on the region, leading to the observation of dense unreacted Li columns. In addition, commercial Li metal foil consists of metallurgic non-uniformity including inclusion impurity, e.g., grain boundary and slip lines (*24*). The surface property variations also affect the wettability and reactivity of different locations on the same Li foil and thus presence of intact Li columns. Three additional locations on cycled Li metal anodes were selected to understand the population of those intact Li columns across the entire Li foils in a later discussion. After extensive cycling, the SEI and reacted Li are entangled forming porous structures in all cycled pouch cells (Fig. S4).

The above discussion reflects the overall response of the entire pouch cell to the external pressures applied vertically on the cell. The change of self-generated pressures are average values detected from the pouch cell upon cycling, which are created through changes of all the layers assembled in the battery. How the pressure is distributed within the surface of the pouch cells and the implications of large-scale electroplating process of Li are still unknown. To investigate this question, a pressure mapping system (see SI for experimental setup) is therefore applied to monitor the pressure distribution on the surface of a 350 Wh/kg pouch cell (Fig. 3 and Fig. S5). After resting for 2 hours, the average pressure on the surfaces of pouch cells is 32.7 psi (Fig. 3A). The pressure detected in the center area of the cell is slightly higher than the rest of the surfaces at open circuit voltage (OCV), suggesting there is still room to further increase the homogeneity of initial pressure applied on the pouch cell.

After the first charge (Fig. 3B), the average pressure on the cell surfaces increased to 81.4 psi due to Li (from NMC) plating on the anode side, creating additional Li layers (and thus pressure) on the 50 µm Li foils as discussed earlier. The pressure distribution at the end of the first charge does not follow the same pattern at OCV and more hotspots are seen in the center (Fig. 3B) which is further amplified at the end of the $5^{th}$ (Fig. 3C) and $20^{th}$ charge (Fig. 3D). At the end of $50^{th}$ (Fig. 3E) and $100^{th}$ (Fig. 3F) charge, the hotspots propagate to the rest regions of cell surfaces suggesting more active areas become available in the cells.

The pressure distribution at the end of discharge (Fig. 3G-3K) follows a similar trend but the amplitude of pressure increase at the end of discharge is much less than that at the end of charge, consistent with our earlier discussions. For example, at the end of the first discharge when the plated Li (from NMC side) goes back to the cathode, the average surface pressure reverses back to 34 psi, close to 32.7 psi at OCV (Fig. 3G). As cycling goes on, the average surface pressure at the end of discharge continues to increase (Fig. 3G-3K) suggesting that after each cycle there are always some byproducts, like SEI accumulated.

At the end of charge, it has been noticed that the amplitude of pressure increase in the center part of pouch cell is consistently much higher than the rest of the regions from the same surface, indicating more reactions are happening in the central area. To understand this phenomenon, the pressure mapping on the cell surface was stopped after more than 100 cycles for post-mortem analysis. A representative Li metal anode from the cycled pouch cell is displayed in Figure 3L. The center part of cycled lithium anode always appears shinier than the four edges (Fig. 3L) in all harvested Li metal foils from the cell. Unreacted dense Li (highlighted in Fig. 3M-3P) is generally seen in all four selected regions from cycled Li (Fig. 3L). However, the population of the intact Li column varies at different locations. The central part of Li anode has much more intact Li (Fig. 3M) than on the edges (Fig. 3N-3P), consistent with the color difference (Fig. 3L). It seems that Li in the center area does not participate much in the electrochemical reactions thus maintaining their original morphologies and metallic shine better than the rest of the same lithium foil. We have checked both sides of the same cycled Li foil (Fig. S6) as well as different Li foil anodes from the same pouch cell (Fig. S7 and S8) and the observation is consistent.

At a first glance, the presence of more intact and shiny Li in the center part is contradictory to the significant increase of pressure in the center of pouch cell (Fig. 3E and 3F) which, as discussed earlier, reflects activity of the materials in that area. However, throughout the cycling, Li utilization is mainly from NMC cathode side. The original Li foil on the anode side only gradually participates in the reaction when needed. The pouch cell used for analysis in Figure 3 is in discharged status meaning the majority of Li from NMC has returned to the cathode side. The reduced utilization of center Li foil indicates that the majority of Li involved in the electrochemical reactions occurring in the central part of cell is from the NMC side. Li stored on the edges of the original Li foil participates more in the cycling compared with the central Li. The pouch cell in Figure 3 was disassembled after 100 cycles. For the pouch cells discussed in Figure 2, even after 300 cycles, all harvested Li foils from the Li metal pouch cells have similar attributes seen in Figure 3L, that is, the center part is much brighter than the rest of Li foils (Fig. S9), confirming that the center Li (of original Li foil) is consumed slower than the edges.

Combining the observation of the drastic pressure increase in the central part of the pouch cell at the end of charge (Fig. 3B-3F), it is hypothesized that during charge Li (from NMC) is not homogeneously plated on the Li foil anode side. Instead, $Li^+$ ions (from NMC) preferably plate in the center of Li foil anode leading to the much faster increase of self-generated pressure in the central area of the pouch cell. Ideally, the electroplating of Li (from NMC) should be almost homogeneous at a slow rate of C/10 charge at least for the first few cycles. However, the slightly higher pressure in the center at OCV shortens the diffusion length of $Li^+$ between cathode and anode and accelerates the electrochemical reactions in the central area which attracts all $Li^+$ nearby towards the center to replenish $Li^+$. Some of the $Li^+$ that are supposed to deposit on the edges can detour to the center of Li anode and are plated there. Similar situations can be found in fast charging Li-ion batteries in which preferred plating of Li happens in the central area of graphite due to pressure variations (*25*).

An anode-free pouch cell is designed to experimentally prove the possibility of $Li^+$ detour in electroplating process. The four edges of copper current collector (anode) are covered by 5 mm wide insulating film (Fig. 4A, see details in Fig. S10) to reduce the anode area so it is smaller than that of the NMC cathode now (Fig. S10). If $Li^+$ from NMC cannot detour when traveling to the anode, the edges of NMC facing the insulating layers of anode will not participate in reactions. Accordingly, the charge and the discharge capacities calculated based on the weight of NMC should decrease because the edges of NMC facing insulating tape cannot be utilized. However, normal capacities are seen during both charge and discharge which nearly overlap with the voltage profiles of a regular cell (without Li foil) in which cathode area is smaller than that of anode, thus confirming that straight travel route may not be necessarily needed for $Li^+$ ions.

The preferred electroplating of Li (from NMC) in the central part of original Li metal anode, driven by the relatively higher pressure applied on the center, explains why the pressure increase is always much faster in the central part of the cell than the other regions (Fig. 3B-3F). Compared to central part of Li foil, less Li (from NMC) is plated on the edges. During subsequent charge, more Li on the edges of Li foil anode can participate in the electrochemical reaction and exposed to the electrolyte, compared to central Li foil which is covered more by Li from the NMC side. While during the first 100 cycles Li from the anode is not used, after repeated cycling, the utilization and thus darkening of Li will happen much earlier on the edges than in the central region of original Li foil. A summary of the hypothesis is plotted in Figure 4B.

Another parallel 350 Wh/kg Li metal pouch cell continues to cycle until > 300 cycles (Fig. S11). Significant pressure increase is captured across the entire cell surface but still with the regions in the center area representing the highest pressure, similar phenomenon area also found for the pressure mapping results of Si-based pouch cells (*26*) although a different pressure fixture is used here. Towards the end of cycling, pressure distribution on the surface of the pouch cell becomes stabilized without much change anymore. The pressure mapping images at the $300^{th}$ and $309^{th}$ cycles (inset of Fig. S11) are almost identical in terms of pressure distribution and amplitudes, confirming that the electrolyte almost completely dries out after 300 cycles. Thus, pressure does not largely impact on the cycling of the dry cell towards the end of cycling.

**Ab Initio Electric Fields:**

To further understand Li$^+$ detour phenomenon during electroplating, we have calculated with ab initio electronic structure methods based on density functional theory, the electrostatic potentials (EP) and electric fields (EF) induced on the neighborhood of the surface of Li metal tablets representing structures at different applied pressures, and thus having different concentrations of Li-nuclei (supplement). We also calculated the EPs and EFs on a larger tablet containing two different nuclei concentrations: a high one around the center of the tablet and a lower one around the peripheral of the tablet (Fig. 4C).

The specific movement of chemical species in any type of material (containing nuclei and electrons) can be studied using the electric field ($\mathscr{E}$), i.e., the force per unit of charge, created internally by the same system components, for example, in a battery during open circuit, charge, and discharge or by external sources, e.g., when charging the battery, in addition to the internal fields. The electric field, mainly, tells the magnitude and direction of the displacements of charged moieties (ions and counterions), as well as the rotational movement of neutral dipolar molecules (such as solvents and additives) to align their electric dipole in the direction of the electric field.

The contribution of electrons to generate electric fields can be obtained solving the Schrödinger Equation for a system of *n* electrons located at points r$_i$ ($r^{(n)}$) and *N* nuclei located at points $R_i$ ($R^{(N)}$), thus,

$$\hat{H}(r^{(n)}, R^{(N)})\Psi(r^{(n)}, R^{(N)}) = E(R^{(N)})\Psi(r^{(n)}, R^{(N)}) \tag{1}$$

In this situation, the array of nuclei positions $R^{(N)}$ is directly determined by the pressure applied to the pouch cell, the more pressure the more concentrated the nuclei (Fig. S12). Solving the above equation with an exact Hamiltonian operator ($\hat{H}$) using ab initio methods such as density functional theory (preferred) or a traditional ab initio (such as HF, MP, CC, CI, etc.), we obtain from this eigenvalue problem the wave function $\Psi$ and the energy $E$. The wavefunction $\Psi$ yields a Slater determinant that can be written as

$$\Psi = det|\psi_i(r_1)\ \psi_i(r_2)\ \psi_i(r_3)\ \cdots\ \psi_i(r_n)| \tag{2}$$

Where the $\psi_i(r_j)$ are the crystal or molecular orbitals. In this shorthand notation of the determinant, the rows of the determinant are created running the dummy index *i* from 1 to *n*. And the wavefunction is a function of 3*n* variables. For the sake of simplicity, we use atomic units and ignore the *n* spin variables and normalization constants, but they are considered in the calculations. Thus, the electron density ($\rho_e$) at any point $r$ can be obtained from summing all occupied orbitals,

$$\rho_e(r) = -\sum_i |\psi_i(r)|^2 \tag{3}$$

And the corresponding nuclei contribution, at points $R_i$ with charges $Z_i$, using the Dirac $\delta$-function can be written as

$$\rho_N(r) = \sum_i \delta(r - R_i)Z_i \tag{4}$$

These two densities give us the electrical potential at any point *r*, *V(r)*,

$$V(r) = \int \frac{\rho_N(r) - \rho_e(r)}{|r-r'|} dr \tag{5}$$

Finally, the electric field $\mathscr{E}$ is obtained from the gradient ($\nabla$) of the potential,

$$\mathscr{E}(r) = -\nabla V(r)$$

For a surface under a non-uniform distribution of local pressures, for example a surface with only two of the pressures/concentrations shown in Figure S12, the net electric field in this non-uniform case can be estimated as the vector sum of their corresponding electric fields, since the interactions have been calculated between elementary particles (nuclei and electrons) so their pair interactions are independent of any other particles around. This electric field is obtained through the formalism explained in Equations 1-5 when applied to the nuclei and electrons of the anodes of different densities that can be associated to different applied pressures. Obviously, for this case of having two different pressures on the same surface, the site with the highest pressure will have the highest concentration of lines of force ($\mathscr{E}$). This reasoning can be extended to a surface with any number of different pressures on its surface. We remark that making a larger anode with a given pressure simply adds the electric fields at the center of the anode surface, regardless of the size of the area, and in all cases the electric fields at the lateral boundaries of the anode surface will be mostly repulsive or of reduced attraction with respect to those in the center of the anode.

To directly demonstrate the effects of a non-uniform pressure, we analyzed a slab having two different concentrations of Li nuclei (Fig. 4C). The full slab has an area of 3L×3L, with 3L = 2.6 nm. The area of the highest concentration site is a square of L×L sitting in the center of the 3L×3L surface. The remainder of the surface has a lower concentration of Li nuclei. Figure 4C shows the lines of electric field (blue) originated by the Li-metal surface. They only represent the direction of the trajectories. The values of the electric field are indicated with an orange background in the most representative places approaching the anode and the electric potential of the black iso-potential contours appear with a green background. The low Li concentration at the left- and right-side surfaces show a repulsive behavior against Li-ions and the central one at high pressure features a much higher attraction, even attracting any Li-ion positioned above the low-pressure sites.

We conclude that the driving forces producing the observed trajectory deflection near the anode surface could be determined or controlled by the pressure applied to the pouch cell, increasing in this way the concentration of Li nuclei in the anode, and triggering further electric field changes that drive more Li-ions near the anode to the high concentration or pressure zones of the anode. These induced electric fields in the neighborhood of the anode are the quantum mechanical effects of the electrons and nuclei charges of the anode.

Thus, ab initio results provide very useful insights to further utilize the external pressure to realize preferred deposition in certain anode designs to eliminate the formation of detrimental Li dendrites that may potentially short the cell internally.

**Conclusions**

This work studies large-scale electrochemical plating of Li$^+$ in realistic 350 Wh/kg Li metal pouch cells under different pressures. It has been discovered that a higher external pressure helps smoothen Li plating which is reflected by the lower self-generated cell pressure due to Li/SEI thickening. However, the appropriately high external pressure extends the cycle life of lithium metal batteries only if there is still sufficient electrolyte in the cell. The homogeneity of electrolyte coverage on large Li metal foils plays a key role in the utilization rate of Li on the Li metal anode side. Mapping of the pressure distribution along pouch cell surface upon cycling reveals that Li$^+$ are preferably plated in the central area of anode side where a relatively high pressure is present. This unique phenomenon of Li$^+$ detour driven by pressure is explained by precise quantum theory calculations and experimentally proved in a pouch cell where cathode area is designed to be larger than anode. This work uncovers very different behaviors of Li$^+$ during large-scale electrochemical plating driven by the pressure and provides new findings critical for developing practically effective solutions to address challenges in rechargeable Li metal batteries.

**Acknowledgements**

This research has been supported by the Assistant Secretary for Energy Efficiency and Renewable Energy, Office of Vehicle Technologies of the U.S. Department of Energy (DOE) through the Advanced Battery Materials Research Program (Battery500 Consortium). The SEM and TEM were conducted in the William R. Wiley Environmental Molecular Sciences Laboratory (EMSL), a national scientific user facility sponsored by DOE's Office of Biological and Environmental Research and located at Pacific Northwest National Laboratory (PNNL). PNNL is operated by Battelle for the DOE under Contract DE-AC05-76RL01830.

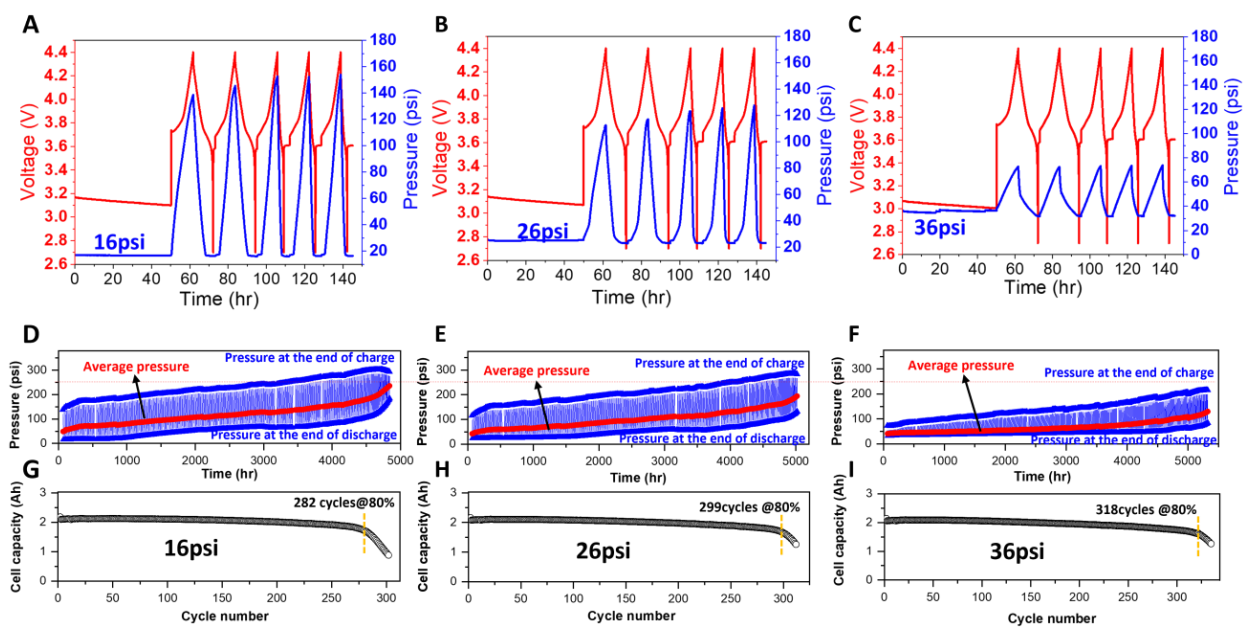

**Figure 1.** *In situ* **pressure monitoring of three 350 Wh/Kg Li/NMC622 pouch cells (2Ah) tested under an initial external pressure of (A, D, G) 16 psi, (B, E, F) 26 psi and (C, F, I) 36 psi. (A-C)** shows the evolution of self-generated pressures during charge/discharge of the three pouch cells in the first five cycles. **(D-F)** compares the pressures detected at the end of charge/discharge during each cycle for the three pouch cells started with different initial pressures. The average pressure is also plotted in (D-F) to compare the amplitude of pressure change generated from three pouch cells upon cycling. **(G-I)** are cycling stability of three Li metal pouch cells tested under different initial pressure of (G) 16 psi, (H) 26 psi and (I) 36 psi. All three pouch cells were charged at 0.1 C and discharged at 0.3 C rates between 2.7 V and 4.4 V.

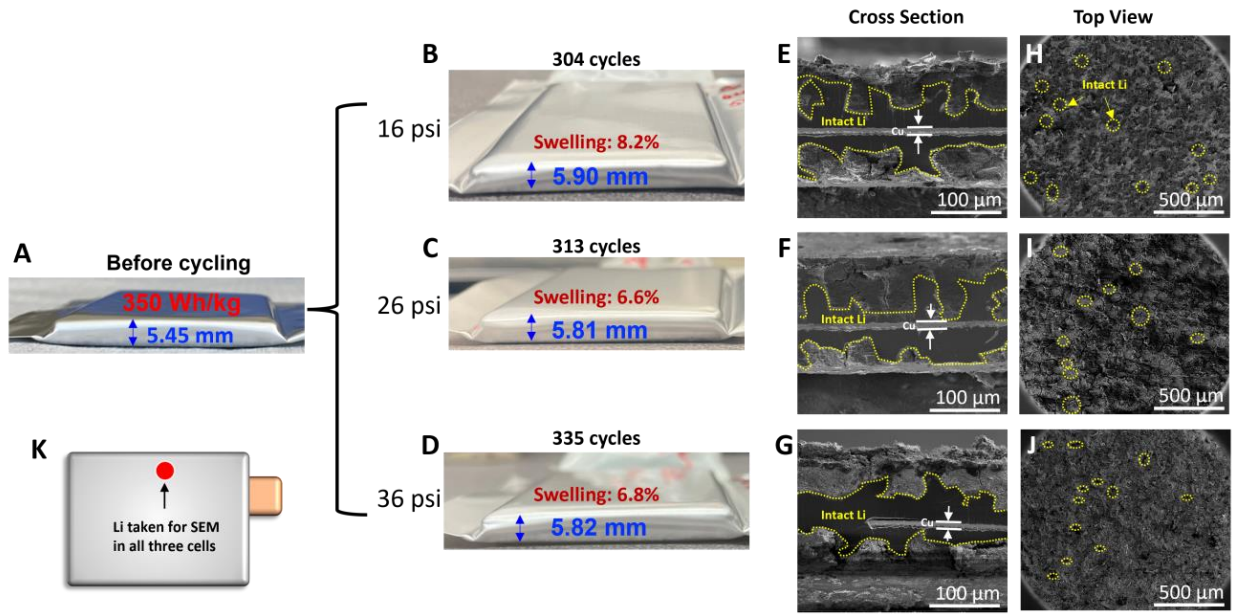

**Figure 2. Li metal pouch cells before and after extensive cycling under different initial pressures applied**. (**A**) As-prepared 350 Wh/kg Li metal pouch cell is about 5.45 mm thick. (**B-D**) Pictures of pouch cells tested under an initial pressure of (**B**) 16 psi after 304 cycles, (**C**) 26 psi after 313 cycles and (**D**) 36 psi after 335 cycles. The swelling rate of all three pouch cells is between 6-8% after extensive cycling. (**E-G**) are the cross section and surface view SEM images of cycled Li metal anode harvested from cycled pouch cells tested at initial pressures of (**E-H**) 16 psi, (**F-I**) 26 psi and (**G-J**) 36 psi. For each pouch cell, the 5[th] Li anode assembled in the cell was taken out for further SEM characterization. On each harvested Li anode, a small piece of cycled Li was cut from the same location on the long edge near copper tab as indicated in (**K**).

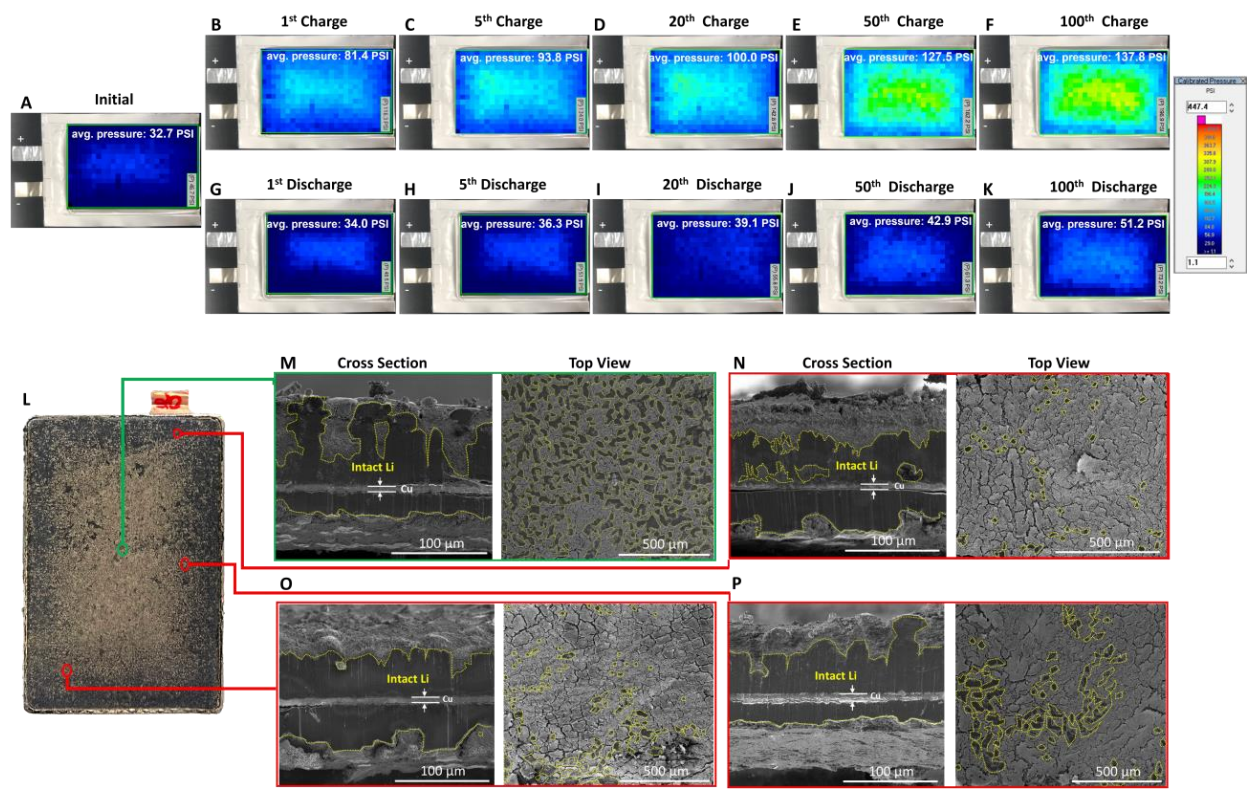

**Figure 3. Mapping of pressure distribution on the surface of a prototype 350 Wh/Kg Li/NMC622 pouch cell (2Ah) during cycling and correlation to lithium metal morphologies after cycling.** **(A)** The average pressure on the surface of the pouch cell is 32.7 psi at OCV status. The pressure applied on the center is slightly higher than the rest area of the pouch cell from the mapping system. Pressure detected on the surface of the pouch cell after the **(B)** 1$^{st}$, **(C)** 5$^{th}$, **(D)** 20$^{th}$, **(E)** 50$^{th}$ and **(F)** 100$^{th}$ charge display continuous increase. The central part of the pouch cell demonstrates faster increase of pressure compared to the rest regions of surfaces. Pressure distribution after the **(G)** 1$^{st}$, **(H)** 5$^{th}$, **(I)** 20$^{th}$, **(J)** 50$^{th}$ and **(K)** 100$^{th}$ discharge. The amplitude of pressure increase on the pouch cell surface is much less at the end of discharge compared to the charge status. **(L)** is the photo of one of the cycled Li metal anodes harvested from cycled Li metal pouch cell used for pressure mapping. The center part of cycled Li metal anode is still metallic shine and is further analyzed by using **(M)** SEM from cross section and top views. Additional three locations from the edges are also selected for SEM characterizations. **(N)** displays the SEM images of Li located on the short edges near copper tab. **(O)** represents the cross-sectional and top view SEM images of Li located in the corner of the cycled Li foil anode far away from the tabbing. **(P)** represents the SEM images of Li located in the long edge of the same Li metal anode as indicated in **(L)**. The pouch cell is in discharge status before being disassembled to harvest the cycled Li metal anode.

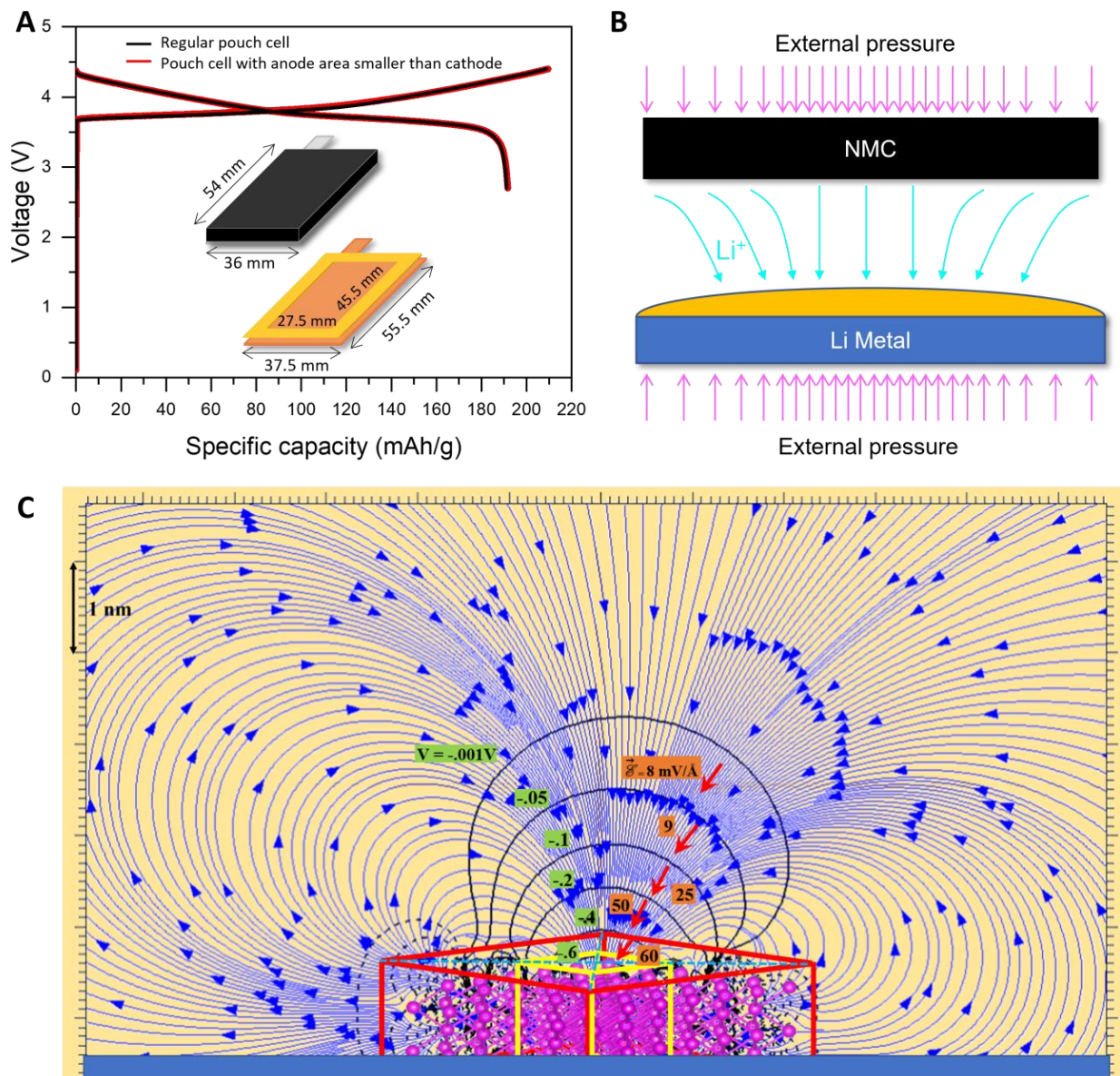

**Figure 4. Experimental and theoretical study of Li⁺ detour behavior during electroplating process.** (**A**) Comparison of the charge and discharge capacity of two anode free pouch cells. One is the regular cell meaning the anode (copper current collector) is larger than NMC cathode, while the other one is the opposite, i.e., the four edges of copper current collector are covered by insulating tape thus the cathode area now becomes larger than that of anode. There is no difference in terms of capacities indicating that Li from NMC side can detour to be plated in the downsized copper current collector. (**B**) Visualization of Li⁺ detour behavior in Li/NMC pouch cell driven by the uneven distribution of externally applied pressure. (**C**) The theoretical calculation of the electric field distribution at Li anode affected by two different external pressures. The electric field lines of force (blue trajectories) are induced by the nuclei and electrons at the anode and interact with the incoming lithium ions as they approach the anode, causing them to deviate from their originally straight paths due to the charging electric field. Note that the regions of highest concentration or pressure of Li nuclei (yellow cube) is the most attractive to lithium ions. Although this seems to be at odds with Coulombic behavior, it is also clear that a higher concentration of nuclei is accompanied by a higher concentration of electrons in the neighborhood. The red parallelepiped (tablet shape) box is the full anode of Li-metal. The central atoms in the yellow parallelepiped are at a higher concentration (pressure) than the reminder Li-nuclei in the red tablet. The electric potential and the electric field are plotted on a vertical plane that cuts diagonally (on the dashed light blue horizonal diagonal) the red and therefore the yellow parallelepipeds. The electric field values are indicated in an orange background and the electric potentials in a green background.